\documentclass[smus]{snow2e}

\begin{document}

\title{
{\normalsize {\rm BUHEP-96-28 \hspace{12.8cm}  October 3, 1996}}
\\ \vspace{-1mm}
{\normalsize {\rm hep-ph/9610240 \hspace{14.9cm} ~}}
\\ \vspace{6mm}
Signals of Dynamical Supersymmetry Breaking in a Hidden Sector $^1$
\\ \vspace{2mm}}

\author{ B. A. Dobrescu $^{(a)}$
%\footnote{dobrescu@budoe.bu.edu} 
and S. Mrenna $^{(b)}$
%\footnote{mrenna@hep.anl.gov} 
 \vspace{2mm} \\
$^{(a)}$ {\normalsize {\it Department of Physics, Boston University,
590 Commonwealth Avenue, Boston, MA 02215, USA }} \\ \vspace{0.5mm}
$^{(b)}$ {\normalsize {\it 
Argonne National Laboratory, 9700 S. Cass Ave.,
Argonne, IL  60439, USA}}
}

\maketitle

%% Get rid of page numbering
\thispagestyle{empty}\pagestyle{empty}

\vspace*{1.05cm}
%vspace{10pt}

\footnotetext[1]{ 
To appear in the ``Proceedings of the 1996 DPF/DPB Summer 
Study on New Directions for High Energy Physics'' (Snowmass 96).}

\begin{abstract}
If supersymmetry is dynamically broken in a hidden sector,
the gauginos typically have unacceptably small masses.
This situation can be corrected if a non-Abelian gauge interaction
becomes strong at a scale of order 1 TeV and induces
dynamical gaugino masses. We discuss the typical signals of this
scenario at present and future colliders.
\end{abstract}

%\section*{Introduction}
\vspace{19pt}
If supersymmetry (SUSY) breaking is communicated from a hidden sector 
to the visible sector by supergravity, then the squarks and sleptons 
acquire masses suppressed by one power of the Planck scale:
$\Lambda_{\rm SB}^2/M_P$ , where $\Lambda_{\rm SB}$ is the SUSY 
breaking scale. Therefore, SUSY is relevant at the
electroweak scale provided $\Lambda_{\rm SB} \sim 10^{11}$ GeV.
The gauginos get masses of order $\Lambda_{\rm SB}(\Lambda_{\rm
  SB}/M_P)^n$,
where $n$ is the lowest dimension of the gauge invariant operators
from the hidden sector \cite{ads}. Current limits on the gluino mass
require $n = 1$, i.e. there should be a gauge singlet in the hidden
sector.
However, the hierarchy between $\Lambda_{\rm SB}$ and $M_P$ requires
dynamical SUSY breaking, and the majority of the models of this type
cannot accommodate gauge singlets \cite{ads,dnns} (possible exceptions
may be found in \cite{singl}). 
Furthermore, mass terms for gauge singlets can be forbidden only by 
global symmetries, which are expected to be violated by Planck scale
effects \cite{planck}. 
Thus, any gauge singlet is likely to have a mass of order
$M_P$, so that a VEV of order $\Lambda_{\rm SB}$ would require an
unacceptable amount of fine tuning.
Note that this problem is more severe than the existence of light
singlets under the Standard Model (SM) gauge group (as the ones used
to generate a $\mu$ term), which may be protected by 
discrete gauge symmetries.

Therefore, hidden sector models may be realistic only if an additional
source of gaugino masses is provided. This can be done, without 
reintroducing a hierarchy problem, if new gauge interactions
become strong at a scale $\Lambda_{\rm G} \sim 1$ TeV. The idea is to
produce dynamical masses for some fields whose spectrum is 
non-supersymmetric due to the usual supergravitational interactions, 
and then to feed
these dynamical masses into one-loop gaugino masses.
A specific example is constructed in ref.~\cite{gluino}.
Here we discuss 
the signatures of this class of models
at present and future colliders. \\
\vfill

\vspace*{1.4cm} 
A striking feature of this scenario, as opposed to the usual
supergravity (SUGRA) 
models which are not concerned with the mechanism for 
dynamical SUSY breaking, is that the gaugino masses
fall off at large momenta. As a result, the radiative corrections
to the squark and slepton masses coming from gauge interactions are
small. This can be seen by evaluating the one-loop gluino
contribution to the squark self energy shown in Fig.~1.

%%%%%%%% FIGURE %%%%%%%%%%%%%%%%%%%%%%%%%%%%%%%%%%%%%%%%
\begin{picture}(200,100)(-10,48)

\thicklines
\put(100,100){\oval(46,40)[t]}
\thinlines
\put(100,100){\oval(46,40)[b]}
\put(100,120){\circle*{14}}

  \put(77,100){\line(-1,0){10}} \put(62,100){\line(-1,0){10}}
  \put(47,100){\line(-1,0){10}} 
  \put(123,100){\line(1,0){10}} \put(138,100){\line(1,0){10}}
  \put(153,100){\line(1,0){10}} 

  \put(74,117){$\tilde{g}$}
  \put(122,117){$\tilde{g}$}
  \put(100,70){$q$}
  \put(35,110){$\tilde{q}$}
  \put(163,110){$\tilde{q}$}

\end{picture}

\vspace{-0.15in}

\makebox[0.5in][l]{Fig.~1.}\makebox[2.5in][l]{\small 
Gluino-quark contribution to the squark mass.}

\makebox[0.5in][l]{}\makebox[2.5in][l]{\small The $\bullet$ is the
  dynamical gluino mass.}
%%%%%%%%%%%%%%%%%%%%%%%%%%%%%%%%%%%%%%%%%%%%%%%%%%%%%%%%%%

\vspace{0.15in}
The quadratic divergence is cancelled by two other 
one-loop diagrams, involving a gluon and the same squark,
respectively. The dynamical gluino mass, represented by a
$\bullet$ in Fig.~1, can be approximated by a momentum 
independent mass, $M_3$, which vanishes at energies above 
$\Lambda_{\rm G}$. Therefore, the one-loop integral should be 
cut-off at $\Lambda_{\rm G}$, and the result is
\begin{equation}
\delta M_{\tilde{q}}^2 \approx \frac{8}{3\pi} 
\alpha_s(\Lambda_{\rm G}) M_3^2
\log\left( \frac{\Lambda_{\rm G}}{M_3}\right) ~,
\end{equation}
where $\alpha_s(\Lambda_{\rm G}) \approx 0.1$ is the QCD
coupling constant at the scale $\Lambda_{\rm G}$.
As we will argue below, the squark mass, $M_{\tilde{q}}$, 
is significantly larger than $M_3$, implying
$\delta M_{\tilde{q}}^2 \ll M_{\tilde{q}}^2$. Similarly, 
the gaugino contributions to the slepton masses are small.

An equivalent description of these radiative corrections
can be given in terms of the renormalization group evolution:
the running of the squark
and slepton masses from $\Lambda_{\rm SB}$ down to $\Lambda_{\rm G}$
is not affected to leading order 
by gauge interactions. Hence, if the Kahler potential
is minimal, then the squarks and sleptons are almost degenerate at the
electroweak scale. The only exceptions are the stops, which are
lighter due to the large radiative corrections induced by the top
Yukawa coupling. 

Since the scale of SUSY breaking in the
spectrum of the fields carrying the new gauge interactions
is of order 1 TeV, we expect the squarks and sleptons
to be rather heavy, in the 0.5 -- 1 TeV range. 
The gluino mass, $M_3$, is suppressed by a factor of
$\alpha_s(\Lambda_{\rm G})$
times some group factor that arises from the coupling to the
fields with dynamical masses. Therefore, $M_3 \sim 100 - 300$
GeV is the typical range.

The Majorana masses for the $U(1)_Y$ gaugino, $M_1$, and $SU(2)_W$
gaugino, $M_2$, 
are proportional to the hypercharge and weak coupling constants
($\alpha_1$ and $\alpha_2$),
respectively, provided that the 
superfields responsible for these masses
transform non--trivially under the SM group.
The usual relations from GUT theories
will not hold in general because the gaugino masses also depend on the
SM charges of the fields carrying the new gauge
interactions. In the model presented in ref.~\cite{gluino}, the 
gaugino mass ratios are given by
\begin{eqnarray}
M_3:M_2:M_1 & = & \frac{1}{3} t(R_3) \alpha_s(\Lambda_{\rm G}) 
 : \frac{1}{2} t(R_2) \alpha_2(\Lambda_{\rm G}) \nonumber\\
 & & :t(R_1) \alpha_1(\Lambda_{\rm G}) ~,
\end{eqnarray}
where $t(R_k)$, $k = 1,2,3$, are the indices of the
color, weak and hypercharge representations 
of the fields with dynamical masses. 
It appears naturally that the inequalities $M_1 < M_2 < M_3$
remain true, with $M_1$ of order 30 GeV and $M_2$ of order
100 GeV. On the other hand, 
if the chiral superfields responsible for the Majorana masses
are weak singlets, which is the most economical alternative,
then $M_2=0$.

\vspace{3pt}
Consider a typical set of parameters in the visible sector:
\begin{eqnarray}
&  \!\!\!
M_1 = 50 \; {\rm GeV}\; , \; M_2 = 100 \; {\rm GeV}\; , \; 
M_3 = 200 \; {\rm GeV}\, ,
&
\nonumber\\
& \!\!\!
\tan\beta=5\; , \; \mu=180 \; {\rm GeV}\; , \; M_{\tilde Q}\simeq
M_{\tilde\ell}  = 800 \; {\rm GeV}\, , 
&
\nonumber\\
& \!\!\!
m_A=900 \; {\rm GeV}\; , \; M_{\tilde t_1}=560 \; {\rm GeV}
\; , \; M_{\tilde t_2}=570 \; {\rm GeV}\, .
&
\label{set1}
\end{eqnarray}
The lightest particles are then the neutralinos 
$(\tilde{N}_1,\tilde{N}_2,\tilde{N}_3,\tilde{N}_4)=(39,75,190,225)$
GeV, charginos $(\tilde{C}_1,\tilde{C}_2)=(69,225)$ GeV,
the gluino ${\tilde g}=249$ GeV, and the Higgs $h$ = 104 GeV.
The largest cross sections at a hadron collider are
$\tilde{C}_1 \tilde{N}_2, \, \tilde{C}_1 \tilde{C}_1$ and 
$\tilde g\tilde g$ production, with 
$\tilde{C}_1 \to W^{*}\tilde{N}_1, \; \tilde{N}_2 \to Z^{*}
\tilde{N_1}$, and $\tilde g\to \tilde{C}_i/\tilde{N}_i
jj$.
The expected signatures are trilepton, dilepton, and 4 jet events
plus missing $E_T$.  LEP2 would be sensitive to chargino pair
production. Despite the fact that the
soft parameters of the model have quite a different origin, the
phenomenology looks very similar to a SUGRA model.  Of course, the
usual relationship between the neutralino/chargino masses and the
gluino mass is modified, giving a possible handle on the underlying
theory.

\vspace{3pt}
An alternative possibility is the set of parameters 
(\ref{set1}) but  with
\begin{equation}
M_2=0 \; , \; \tan\beta=1.5 \; , \; \mu=-50 \; {\rm GeV} ~.
\end{equation}
This corresponds to the
case when the fields responsible for the Majorana masses are
SU(2)$_W$ singlets.  The values of
$\tan\beta$ and $\mu$ are chosen to give
compatibility with LEP limits.  
The mass spectrum is then 
$(\tilde{N}_1,\tilde{N}_2,\tilde{N}_3,\tilde{N}_4)=(31,47,88,115)$
GeV, 
$(\tilde{C}_1,\tilde{C}_2)=(53,112)$ GeV,
${\tilde g}=249$ GeV, and $h$ = 73 GeV.
The largest cross sections at a hadron collider are
$\tilde{N}_1 \tilde{N}_1, \, \tilde{N}_1 \tilde{N}_2, \, 
\tilde{C}_1 \tilde{C}_1, \, \tilde{N}_1 \tilde{C}_1$ and 
$\tilde g\tilde g$ production.
As before, $\tilde{C}_1 \to W^{*}\tilde{N}_1$, but there is a sizable 
branching
fraction for $\tilde{N}_2 \to \gamma \tilde{N}_1$, giving a possible
signal of
$\gamma$ and missing $E_T$.  The usual trilepton signature would
be absent.  LEP2 would be sensitive to chargino and neutralino
pair production, the latter giving a striking $\gamma$ plus missing
$E$ signal.

So far we have seen that the potential
signals of hidden sector dynamical SUSY breaking 
at the existing colliders and their upgraded versions
are related only to the gaugino and Higgs sectors, and 
are peculiar only for a class of models. However, if squark and
slepton production will be possible at the next
generation of colliders, then the spectrum of these particles
will provide an important test of hidden sector dynamical SUSY
breaking. The reason is that in most models of
SUSY breaking \cite{ssb} the squarks are significantly heavier than
the sleptons, while the models discussed here predict 
approximate squark-slepton degeneracy.

At even higher energies it will be possible to probe the
strongly coupled sector responsible for gaugino masses.
If pseudo Goldstone bosons result from the
strong dynamics, then they will be the first particles beyond the
minimal supersymmetric SM to be seen. In the model discussed in
\cite{gluino}
the chiral symmetry of the fields carrying the new strong interaction
is spontaneously broken leading to three Goldstone bosons.
This chiral symmetry is also explicitly broken such that the Goldstone
bosons get masses of order 1 TeV.
Since the new non-Abelian gauge interactions become strong at a scale
of order 1 TeV, scaling from QCD suggests that the lightest composite
states beyond the Goldstone bosons will have masses in the 5 -- 10 TeV
range.
In conclusion, if SUSY is dynamically broken in a hidden sector,
a very rich phenomenology will emerge at future colliders.

\vspace{5pt}
%%%%%%%%%%%%%%%%%%%%%%%%%%%%%%%%%%%%%%%%%%%%%%%%%%%%%%%%%%%%%%%
$ Acknowledgements.$
We would like to thank Sekhar Chivukula and Joseph Lykken for
useful discussions.
This work was supported in part by the National Science
  Foundation under grant PHY-9057173, and by the Department of Energy
  under grant DE-FG02-91ER40676.

%%%%%%%%%%%%%%%%%%%%%%%%%%%%%%%%%%%%%%%%%%%%%%%%%%%%%%%%%%%%%%%

\end{document}